\documentclass{svproc}
\usepackage{url}

\usepackage{amsmath}
\usepackage{bbold}
\usepackage{xcolor}
\usepackage{graphicx}
\usepackage{booktabs}
\usepackage{multirow}

\usepackage{caption}
\usepackage{float}

\usepackage{listings}

\usepackage[colorlinks=true, linkcolor=blue, urlcolor=blue, citecolor=blue]{hyperref}

\begin{document}
\mainmatter              
\title{Unlocking LLM Code Correction\\ with Iterative Feedback Loops}
\titlerunning{Iterative Feedback Loops}  
%
\author{Le Zhang \and Suresh Kothari}
\authorrunning{Le Zhang et al.} 
%
\tocauthor{Le Zhang, Suresh Kothari}
\institute{Iowa State University, Ames Iowa 50010, USA\\
\email{\{lezhang, kothari\}@iastate.edu}}

\maketitle              

\begin{abstract}
Large Language Models have shown remarkable capabilities in code generation. However, most existing evaluations focus only on single-attempt accuracy and overlook the iterative refinement process that is central to real-world programming. This study presents a systematic investigation of LLMs’ ability to rectify their own code through execution feedback. Using real-world programming problems across four models and two major programming languages, this study evaluates performance using iterative refinement framework where LLMs receive compiler error messages and testcase feedback after each attempt. This study introduces metrics to evaluate code failures, analyze rectification patterns, and compare the effectiveness of reasoning and non-reasoning models, offering actionable insights into both the understanding and practical application of feedback loops in LLM-driven code generation systems. Results show that reasoning models consistently improve over iterations, substantially outperforming non-reasoning models in leveraging feedback, while syntactic and runtime errors are far more tractable than logical or algorithmic failures.

\keywords{Large Language Models (LLMs), Code Generation, Feedback Loop, Iterative Refinement, Model Evaluation}
\end{abstract}
\section{Introduction}\label{sec:introduction}

Large Language Models (LLMs), such as DeepSeek and ChatGPT, are capable of generating code from natural language specifications and have the potential to significantly reduce human effort in software development. However, these efficiency gains can quickly diminish if the generated code fails during testing, necessitating human intervention for correction. Therefore, it is crucial that LLMs not only generate but also rectify the code. Analogous to human programmers, LLMs require feedback loops to iteratively improve their code. In such loops, testcase failures and error messages are fed back to the model, which in turn attempts to correct the code. Despite its importance, there is limited knowledge about the effectiveness of feedback loops in LLMs, and there are no established practical guidelines for their implementation.

This study aims to systematically investigate LLMs with feedback loops by experimenting with state-of-the-art models. While previous work has extensively evaluated LLM code generation \cite{austin2021program,chen2021evaluating,nijkamp2023codegen,wang2023review,jiang2023selfevolve,wang2023chatcoder,butt2024codeit,jiang2024survey,yu2024outcome,zhong2024debug,chen2025pass,zhang2025holistic}, these studies primarily focus on metrics such as pass@1 or pass@k and do not explore the role of feedback loops in code rectification.

This study focus on aspects that are particularly relevant to real-world software development, where feedback loops typically operate automatically. In this study, researchers implement an automated feedback loop consisting of the following steps: (i) execute the code generated by the LLM on a test case, (ii) construct a prompt that incorporates any resulting failure messages, and (iii) provide this prompt to the LLM as execution feedback. The loop initiates with code that has failed during testing. For each programming task, researchers employ several hundred test cases designed to cover corner cases, logical errors, and violations of space and time constraints. The feedback loop iterates until the LLM either successfully rectifies the code or exhausts the opportunity for improvement.

Conducting a systematic investigation of feedback in LLMs entails several challenges. First, it requires establishing a baseline understanding of LLM failures across diverse scenarios, including logical errors and violations of computational constraints. Second, realistic feedback loops must emulate practical software development, comprehensive evidence of failures while allowing for code rectification. Third, careful monitoring of the feedback process is necessary to track and characterize the types of improvements achieved and the failures that persist. Finally, the study must account for variability across LLMs, including differences between reasoning and non-reasoning models, to assess whether state-of-the-art reasoning models consistently outperform their non-reasoning counterparts.

The main contributions of this research are threefold. First, a novel methodology for systematically exploring and evaluating the impact of feedback in LLMs is proposed. Second, an extensive experimental setup that enables a rigorous investigation of LLM code rectification in realistic conditions is developed. Third, actionable insights for software development, including metrics to assess code failures, patterns of rectification, and the comparative effectiveness of reasoning versus non-reasoning models are provided. Collectively, these contributions advance both the understanding and practical application of feedback loops in LLM-driven code generation.

The remainder of this paper is organized as follows. Section~\ref{sec:research_methodology} details the research methodology. Section~\ref{sec:results} presents the experimental results, followed by Section~\ref{sec:discussion}, which discusses notable cases and the limitations of this study. Section~\ref{sec:related_work} reviews existing research. Finally, Section~\ref{sec:conclusion} summarizes the key findings and outlines directions for future research.
\section{Research Methodology}\label{sec:research_methodology}

\subsection{Research Questions} 

This study investigates how LLMs revise code using execution feedback. To explore this challenging problem, this study is framed through the following research questions.

\vspace{0.08in}
\noindent\textbf{RQ1}: How do LLMs perform across different problem difficulties using standard pass@1 evaluation metrics? 

\textit{This question establishes the baseline capabilities of LLMs, revealing how model performance scales with task complexity and where failures arise. Understanding these baseline limitations, across functional errors, unmet constraints, and efficiency bottlenecks, is crucial for assessing the potential impact of feedback-driven code rectification. Detailed analysis and findings are presented in Section~\ref{sec:result_rq1}.}

\vspace{0.08in}
\noindent\textbf{RQ2}: To what extent do LLMs respond to instructive prompts for algorithmic optimization?

\textit{This question investigates whether explicit guidance improves the efficiency of generated code. Responsiveness to such prompts reflects an LLM’s ability to leverage execution feedback, a prerequisite for effective iterative refinement. Results are presented in Section~\ref{sec:result_rq2}.}

\vspace{0.08in}
\noindent\textbf{RQ3}: How effectively can LLMs use execution feedback to correct code with errors?

\textit{This question examines the ability of LLMs to iteratively correct code that fails on the first attempt. Assessing this capability reveals the potential for autonomous code improvement and informs practical considerations for using iterative feedback loops. Detailed results are presented in Section~\ref{sec:result_rq3}.}

\vspace{0.08in}
\noindent\textbf{RQ4}: How does iterative refinement performance differ between reasoning and non-reasoning LLMs?

\textit{This question compares reasoning and non-reasoning models in their iterative refinement performance, highlighting factors that influence self-correction and guiding model selection for feedback-driven tasks. Results and analysis are presented in Section~\ref{sec:result_rq4}.}

\vspace{0.08in}
\noindent\textbf{RQ5}: Which types of errors, syntactic, logical, or algorithmic, are most effectively corrected by iterative refinement?

\textit{This question analyzes which error types are most amenable to correction, comparing recovery rates across models, programming languages, and error categories. Understanding these patterns helps systematically map the strengths and limitations of iterative feedback-driven code generation. Results are presented in Section~\ref{sec:result_rq5}.}

\subsection{Datasets and Models}

A controlled experiment is conducted to evaluate LLM performance in code generation, systematically assessing models across programming languages and establishing a baseline for subsequent analysis.

\begin{table}[H]
\caption{Problem distribution by difficulty tiers.}
\label{tab:datasets}
\begin{center}
\begin{small}
\begin{sc}
\begin{tabular}{lccc@{\hspace{1em}}c}
\toprule
Dataset & Easy & Medium & Hard & Total \\
\midrule
Core Dataset & 150 & 150 & 150 & 450 \\
Strain Dataset & 0 & 101 & 99 & 200 \\
Challenge Dataset & 0 & 6 & 26 & 32 \\
\bottomrule
\end{tabular}
\end{sc}
\end{small}
\end{center}
\vskip -0.2in
\end{table}

\subsubsection{Data Collection}\label{sec:datasets}

Three datasets were used, each serving a distinct purpose:

\begin{itemize}
    \item \textbf{Core Dataset}: 450 randomly selected LeetCode problems evenly distributed across difficulty levels (150 easy, 150 medium, 150 hard), covering algorithms and data structures such as greedy algorithms, sorting, binary search, and tree-based problems. This dataset provides a representative benchmark for general code generation quality.

    \item \textbf{Strain Dataset}: A 200-problem subset of the Core Dataset manually selected for algorithmic efficiency based on size of the inputs, emphasizing optimized implementations that meet strict time constraints. This dataset evaluates models' ability to produce computationally efficient code.

    \item \textbf{Challenge Dataset}: Comprising the 32 most frequently failed problems across all models and languages in the baseline evaluation. These problems are rarely solved on the first attempt, making this dataset ideal for assessing iterative refinement and feedback-driven improvement.
\end{itemize}

LeetCode was chosen for its structured repository, rich test cases, clear categorization, and emphasis on time and space efficiency, which aligns with the objectives of evaluating algorithmic performance and execution efficiency. Table~\ref{tab:datasets} summarizes the key characteristics of these datasets.

\begin{table}[H]
\caption{Models used in our study.}
\label{tab:models}
\begin{center}
\begin{small}
\begin{sc}
\begin{tabular}{l@{\hspace{2em}}cccc}
\toprule
Model & \begin{tabular}{@{}c@{}}Size\\ (B)\end{tabular} & \begin{tabular}{@{}c@{}}Release\\ Year\end{tabular} & \begin{tabular}{@{}c@{}}Open\\ Source\end{tabular} & \begin{tabular}{@{}c@{}}Reasoning\\ Model\end{tabular} \\
\midrule
DeepSeek-R1 & 671 & 2025 & $\surd$ & $\surd$ \\
DeepSeek-V3 & 671 & 2024 & $\surd$ & $\times$ \\
GPT-o4-mini & 8 (Est.) & 2025 & $\times$ & $\surd$ \\
GPT-4.1-mini & 8 (Est.) & 2025 & $\times$ & $\times$ \\
\bottomrule
\end{tabular}
\end{sc}
\end{small}
\end{center}
\vskip -0.2in
\end{table}

\subsubsection{Selection of Models:}\label{sec:models}

Four state-of-the-art LLMs are evaluated: DeepSeek-R1 and DeepSeek-V3 (DeepSeek), and GPT-o4-mini and GPT-4.1-mini (OpenAI). These models were selected for accessibility, popularity, and representation of state-of-the-art code generation approaches (Table~\ref{tab:models}). By comparing DeepSeek-R1 vs. DeepSeek-V3 and GPT-o4-mini vs. GPT-4.1-mini, the influence of reasoning capabilities on coding and iterative refinement performance can be isolated.

\subsection{Experimental Design: Baseline Experiment}\label{sec:baseline_design}

While baseline $pass@1$ experiments measure single-attempt performance, real-world coding often involves iterative refinement based on compiler or runtime feedback. To evaluate LLMs' ability to improve solutions over multiple attempts, iterative refinement experiments are conducted where models leveraged testcase results and execution feedback as a form of \emph{instructive guidance}.

\subsubsection{Parameter Configuration}\label{sec:baseline_parameter}

To ensure consistent outputs, key decoding parameters were carefully configured. GPT-o4-mini does not allow user-configurable parameters and was evaluated using its defaults. For all other models, we set \textbf{top-p = 0.95}, following prior studies that show this value improves $pass@1$ performance~\cite{holtzman2019curious,zheng2024gpt,JanSiml,coignion2024performance,deepseer1_2025}. This setting broadens the sampling distribution, increasing the likelihood of generating correct solutions.

Temperature controls token randomness. While higher values may help multi-attempt metrics like $pass@k$, they can reduce single-attempt ($pass@1$) performance~\cite{zheng2024gpt}. We set \textbf{temperature = 0.1} to balance output diversity and determinism, ensuring stable, reproducible results.

\subsubsection{Experimental Procedure}

Four LLMs were tasked with solving 450 problems from the Core Dataset in both Python and Java. Each model had a single attempt per problem per language. Generated code was submitted to LeetCode via a custom API, evaluated using official test cases, and results collected in JSON format. All experiments adhered to LeetCode's usage policies to ensure ethical compliance and minimal service disruption. Performance was assessed using the $pass@1$ metric (Section~\ref{sec:metrics}).

\subsubsection{Prompt Design}\label{sec:baseline_prompt}

A structured prompt was used to ensure consistent and high-quality outputs (Figure~\ref{fig:prompt_baseline}). The prompt instructs the model to act as ``a software developer,'' specifies the target programming language, includes the full problem statement with examples and constraints, and provides a partially completed code snippet along with sample test cases. The prompt concludes by requesting only the final code solution, without explanations, to facilitate automated evaluation.

This design minimizes ambiguity, standardizes outputs across models, and promotes functional correctness and computational efficiency, aligning with real-world software development practices.

\begin{figure}[!t]
\centering
\fbox{
    \begin{minipage}{0.9\textwidth}
    \textcolor{cyan}{\# Start of the Prompt}\\
    \textcolor{orange}{You are a \textbf{software developer}. Implement a solution in \textbf{Python} for the following coding problem.}\\
    
    \textbf{Problem Description:}\\
    \textcolor{black}{Given an integer $num$, repeatedly add all its digits until the result has only one digit, and return it.}\\
    
    \textbf{Example:}\\
    Input: $num = 38$\\
    Output: 2\\
    Explanation: The process is\\
    $38 \rightarrow 3 + 8 \rightarrow 11$, then $11 \rightarrow 1 + 1 \rightarrow 2$\\
    Since 2 has only one digit, return it.\\
    
    \textbf{Constraints:}\\
    $0 \leq num \leq 2^{31} - 1$\\

    \textbf{Code Snippet:}\\
    class Solution:
    
    \qquad def addDigits(self, num: int) $\rightarrow$ int:\\

    \textbf{Testcases:}\\
    $num=38$; $num=0$\\

    \textbf{Additional Instructions:}\\
    \textcolor{brown}{Follow the input constraints and write your code starting from the given code snippet. Ensure the code is well-formatted and adheres to best practices. Write the executable code only, avoid unnecessary explanations or comments.}
    
    \textcolor{cyan}{\# End of the Prompt}
    \end{minipage}
}
\caption{Prompt Structure Example for Baseline Experiment}
\label{fig:prompt_baseline}
\vspace{-0.1in}
\end{figure}

\subsection{Experimental Design: Iterative Refinement Experiment}

While the baseline $pass@1$ experiments provide a measure of single-attempt performance, real-world coding often involves iterative refinement based on execution feedback. To evaluate LLMs' ability to improve solutions over multiple attempts, iterative refinement experiments are conducted using test results and runtime feedback.

\vspace{-0.1in}
\subsubsection{Parameter Configuration}

GPT-o4-mini was evaluated using default parameters, as they are not user-configurable. For the other three models, we set \textbf{temperature = 0.9}, following prior work showing higher temperatures improve multi-attempt code generation~\cite{zheng2024gpt}.

Since top-p guidance for iterative tasks is limited, a preliminary calibration is conducted using DeepSeek-R1 over 10 iterations on a subset of problems. Performance peaked at \textbf{top-p = 0.3}, which was applied to all models in iterative experiments (Table~\ref{tab:iterative_top-p_calib}).

\begin{table}[H]
\vskip -0.1in
\caption{Metrics by different top-p values.}
\label{tab:iterative_top-p_calib}
\begin{center}
\begin{small}
\begin{sc}
\begin{tabular}{l@{\hspace{1em}}c@{\hspace{1em}}c@{\hspace{1em}}c@{\hspace{1em}}c@{\hspace{1em}}c}
\toprule
Metric & 0.1 & 0.3 & 0.5 & 0.7 & 0.9 \\
\midrule
ISR@10 & 65.6\% & \textbf{68.8\%} & 68.8\% & 65.6\% & 62.5\% \\
MIS & 5 & \textbf{4} & 6 & 5.5 & 4 \\
\bottomrule
\end{tabular}
\end{sc}
\end{small}
\end{center}
\vskip -0.2in
\end{table}

\subsubsection{Iteration Limit Setup}\label{sec:max-iter}

\begin{figure}[!t]
\centering
\includegraphics[width=0.75\linewidth]{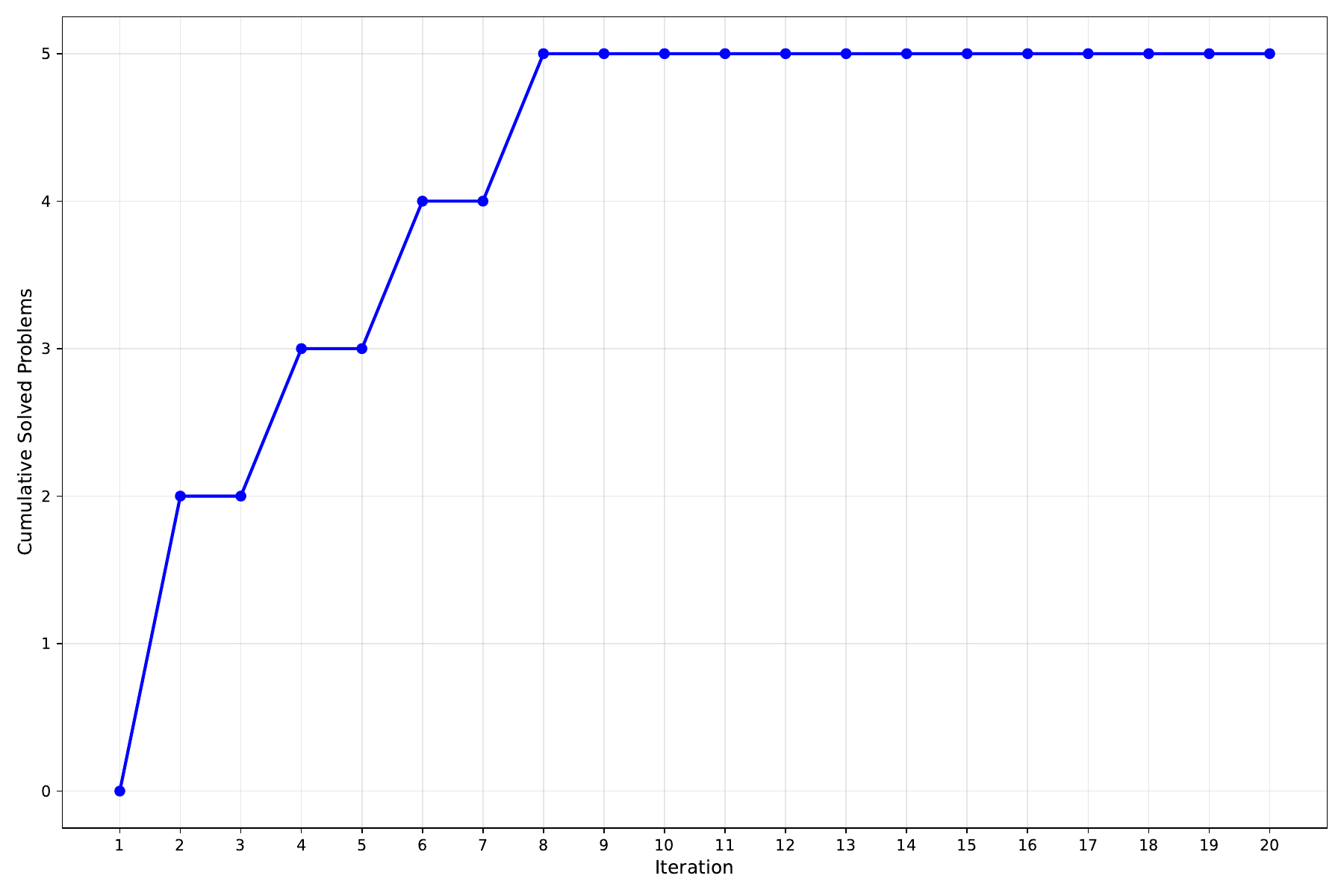}
\caption{Iteration limit calibration with DeepSeek-R1.}
\label{fig:max-iter}
\vskip -0.in
\end{figure}

To balance performance and computational cost, we determined an appropriate iteration limit via a preliminary experiment on 10 randomly sampled problems from the Challenge Dataset. Figure \ref{fig:max-iter} showed that success rates plateaued after the 8th iteration. Accounting for marginal cases, we set a maximum of \textbf{10 iterations} per problem.

\vspace{-0.1in}
\subsubsection{Experimental Procedure:}\label{sec:iterative_procedure} 

\begin{figure}[!t]
\centering
\includegraphics[width=0.95\linewidth]{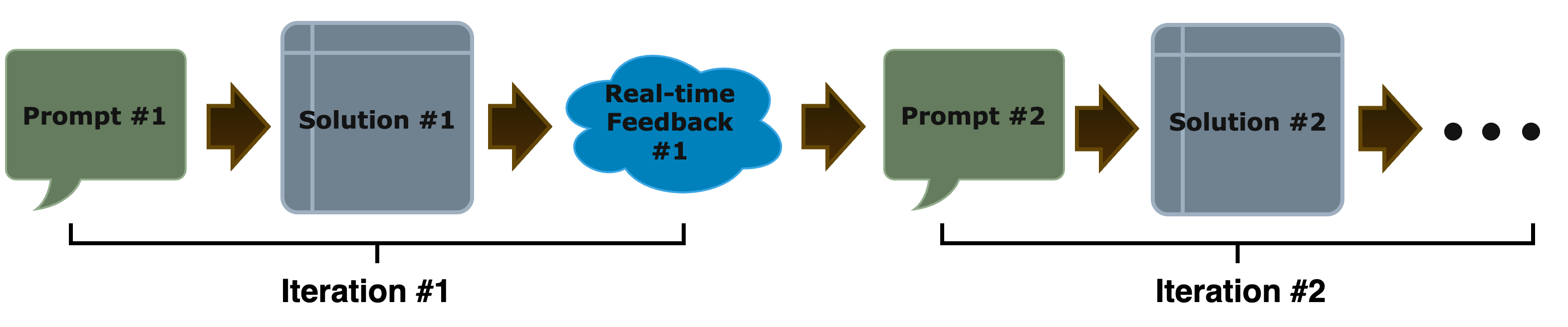}
\caption{Iterative Experiment Procedure}
\label{fig:iter-procedure}
\vskip -0.1in
\end{figure}

As demonstrated in Figure \ref{fig:iter-procedure}, the iterative framework uses a multi-turn mechanism: each iteration involves (1) sending a prompt, (2) generating a solution, and (3) receiving \emph{execution feedback}. The initial prompt follows the baseline structure (Section~\ref{sec:baseline_prompt}). Subsequent prompts include all previous solutions and feedback, giving the model full context. Iterations continue until a correct solution is obtained or the 10th iteration is reached. All solutions and results were stored for analysis using metrics from Section~\ref{sec:metrics}.

\begin{figure}[!t]
\centering
\fbox{
    \begin{minipage}{0.95\textwidth}
    \# \textbf{Compile Error Feedback}:\\
    Your code crashed at a compile error: [\textit{Error Message}]\\
    Fix your code with above information.\\

    \# \textbf{Runtime Error Feedback}:\\
    Your code crashed at a runtime error: [\textit{Error Message}]\\
    Fix your code with above information.\\

    \# \textbf{Wrong Answer Feedback}:\\
    Your code generated wrong outputs at testcase: [\textit{Testcase}]\\
    Expected output: [\textit{Expected Output}]\\
    Actual output: [\textit{Actual Output}]\\

    \# \textbf{Time Limit Exceeded Feedback}:\\
    Your code exceeded the maximum runtime allowance.\\
    Optimize the time complexity of your algorithm.\\

    \# \textbf{Memory Limit Exceeded Feedback}:\\
    Your code exceeded the maximum memory allowance.\\
    Optimize the space complexity of your algorithm to reduce memory usage.

    \end{minipage}
}
\caption{Examples of execution feedback}
\label{fig:feedback_design}
\vspace{0.1in}
\end{figure}

\begin{figure}[!t]
\centering
\fbox{
    \begin{minipage}{0.95\textwidth}
    [
    
    \quad \{ role: "user", content: "You are a software developer. Implement ..." \},
    
    \quad \{ role: "assistant", content: "Solution code \#1"\},
    
    \quad \{ role: "user", content: "Your code crashed at a compile error: ..."\},
    
    \quad \{ role: "assistant", content: "Solution code \#2"\},

    \quad \{ role: "user", content: "Your code generated wrong output ..."\},
    
    \quad \{ role: "assistant", content: "Solution code \#3"\},
    
    \qquad ...

    \quad \{ role: "user", content: "Your code exceeded the maximum memory ..."\},

    \quad \{ role: "assistant", content: "Solution code \#10"\}
    
    ]
    \end{minipage}
}
\caption{Example of multi-turn conversation with LLM}
\label{fig:prompt_iter}
\vspace{0.1in}
\end{figure}

\vspace{-0.1in}
\subsubsection{Execution Feedback}

LeetCode categorizes results as Accepted, Compile Error, Runtime Error, Wrong Answer, Time Limit Exceeded, and Memory Limit Exceeded. Each error type can typically be corrected in specific ways (e.g., Time Limit Exceeded via algorithmic optimization). \textbf{Execution feedback} is provided to guide LLMs toward effective corrections, framing the feedback as actionable guidance in the form of instructive prompts (Figure~\ref{fig:feedback_design}).

\subsubsection{Prompt Design}

A standard multi-turn JSON-based prompt format is adopted (Figure~\ref{fig:prompt_iter}). Prompts are labeled as ``user'' and model responses as ``assistant''. In each iteration, the LLM accesses the full conversation history, enabling it to leverage prior solutions and \emph{instructive feedback} to generate improved outputs over successive iterations.

\vspace{0.2in}
\subsection{Evaluation Metrics}\label{sec:metrics}

\subsubsection{pass@1 Metric}
In the baseline experiment, each model had a single attempt to solve every problem in each programming language. Generated solutions were submitted to LeetCode, and the results recorded. For a dataset of $N$ problems, let $S_i = 1$ if the model passes problem $i$ on its first attempt, and $S_i = 0$ otherwise. The \textbf{pass@1} score is then computed as:

\begin{equation}
\text{pass@1} = \frac{1}{N} \sum_{i=1}^{N} S_i
\label{eq:pass@1}
\end{equation}

where $\text{pass}@1 \in [0, 1]$ represents the proportion of problems successfully solved on the first attempt.

\vspace{0.1in}
\subsubsection{ISR@k Metric:}
In the iterative refinement framework, each LLM is allowed up to $k$ iterations per problem. A problem is marked as a PASS if any iteration produces a correct solution; otherwise, it is marked as a FAIL. We define the \textbf{Iterative Success Rate (ISR@k)} as the percentage of problems solved within $k$ iterations:

\begin{equation}
\text{ISR@}k = \frac{1}{N} \sum_{i=1}^{N} \mathbb{1} \left( \max_{j=1,\dots,k} S_{i}^{(j)} = 1 \right)
\end{equation}

where:
\begin{itemize}
    \item $N$ is the total number of problems,
    \item $S_i^{(j)} = 1$ if iteration $j$ on problem $i$ passes all test cases (otherwise 0),
    \item $\mathbb{1}(\cdot)$ is the indicator function, returning 1 if the condition is true and 0 if false.
\end{itemize}

\vspace{0.1in}
\subsubsection{MIS Metric:}
Consider a scenario in which two LLMs achieve the same ISR@k value. Model-A may solve most problems within three iterations, whereas Model-B requires more than eight iterations for the same problems. Clearly, Model-A is more efficient, but ISR@k alone cannot capture this distinction.

To address this, we introduce the **Median Iterations to Solve (MIS)** metric, which quantifies the typical number of iterations a model requires to solve problems and highlights differences in iteration efficiency.

For a dataset of (N) problems, each problem (p) is assigned a score $(S_p)$ as follows:

\begin{itemize}
\item If problem (p) is solved at iteration (i) $((i \le k))$, set $(S_p = i)$.
\item If problem (p) remains unsolved after (k) iterations, set $(S_p = k+1)$ to distinguish it from problems solved on the last iteration.
\end{itemize}

The MIS is then defined as the median of these scores:
\begin{equation}
\text{MIS} = \text{median}(S_1, S_2, \dots, S_N)
\end{equation}

Lower MIS values indicate higher efficiency, reflecting a model's ability to reach correct solutions in fewer iterations.

\subsection{Replication Package}
All artifacts related to this study, including LeetCode APIs and tools, LeetCode problem datasets, experimental results, and LLM-generated code are available in this public repository: \href{https://github.com/lezhangisu/LLM-Code-Correction}{https://github.com/lezhangisu/LLM-Code-Correction}.

\section{Experimental Results and Data Analysis}\label{sec:results}

\subsection{RQ1: Baseline Performance}\label{sec:result_rq1}

\begin{table}[H]
\caption{pass@1 score by difficulty tier and programming language.}
\label{tab:baseline_results}
\begin{center}
\begin{small}
\begin{sc}
\begin{tabular}{lcc@{\hspace{1em}}c@{\hspace{1em}}c@{\hspace{1em}}c@{\hspace{1em}}c}
\toprule
\begin{tabular}{@{}c@{}}Difficulty\\ Tier\end{tabular} & \begin{tabular}{@{}c@{}}Programming\\ Language\end{tabular} & \begin{tabular}{@{}c@{}}DeepSeek\\ V3\end{tabular} & \begin{tabular}{@{}c@{}}DeepSeek\\ R1\end{tabular} & \begin{tabular}{@{}c@{}}GPT\\ 4.1-mini\end{tabular} & \begin{tabular}{@{}c@{}}GPT\\ o4-mini\end{tabular} \\
\midrule
\multirow{2}{*}{Easy} 
& Python & 99.33 & \textbf{100.00} & 97.33 & 98.67 \\
& Java & \textbf{98.67} & \textbf{98.67} & 97.33 & \textbf{98.67} \\
\midrule
\multirow{2}{*}{Medium}
& Python & 71.33 & 86.67 & 76.67 & \textbf{88.67} \\
& Java & 70.67 & 86.00 & 74.00 & \textbf{89.33} \\
\midrule
\multirow{2}{*}{Hard}
& Python & 46.67 & 65.33 & 54.67 & \textbf{80.00} \\
& Java & 45.33 & 62.67 & 54.00 & \textbf{74.00} \\
\midrule
\multirow{2}{*}{Overall}
& Python & 72.44 & 84.00 & 76.22 & \textbf{89.11} \\
& Java & 71.56 & 82.44 & 75.11 & \textbf{87.33} \\
\bottomrule
\end{tabular}
\end{sc}
\end{small}
\end{center}
\vskip -0.1in
\end{table}

Following the experimental procedure outlined in Section \ref{sec:baseline_design}, we evaluated the coding capabilities of four LLMs across two programming languages using the pass@1 metric. The results, summarized in Table \ref{tab:baseline_results}, provide a baseline view of how well these models perform across languages and difficulty levels.

Overall, all models achieved high accuracy on easy problems, but performance declined notably as problem difficulty increased. DeepSeek-V3 and GPT-4.1-mini exhibited the steepest drops on medium and hard tasks, whereas GPT-o4-mini maintained the most stable and highest overall performance. DeepSeek-R1 performed moderately across all levels, suggesting a balanced but less specialized capability profile.

When comparing languages, Python consistently yielded higher pass@1 scores than Java across nearly all models and difficulty tiers. This trend indicates that LLMs are generally more proficient in Python-based code generation, likely due to Python's simpler syntax\cite{lappi2023replication}, dynamic typing\cite{vitousek2016gradual}, and greater representation in public training data\cite{Cass_2025}.

\begin{figure}[!t]
\centering
\includegraphics[width=0.8\linewidth]{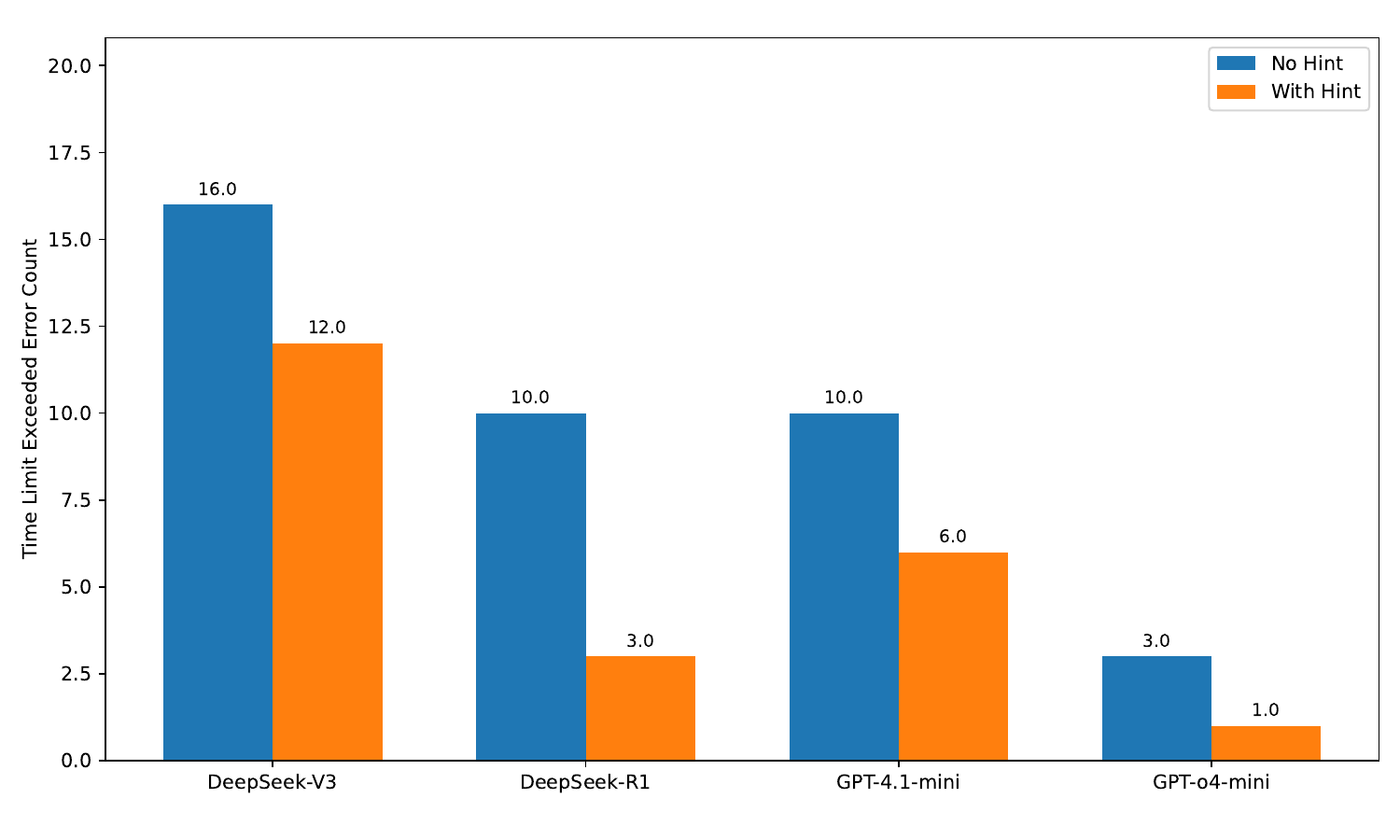}
\caption{Time Limit Exceeded error count with and without suggestive hint.}
\label{fig:baseline_hint}
\vspace{-0.1in}
\end{figure}

\subsection{RQ2: Prompt Sensitivity Analysis}\label{sec:result_rq2}

Building on the baseline results, we observed that while LLMs performed well on many problems, they still struggled to solve more complex tasks—particularly those requiring algorithmic optimization. To examine whether instructive guidance could mitigate this limitation, we re-evaluated all models on 200 problems from the Strain Dataset (Section~\ref{sec:datasets}). In this experiment, we appended a one-line instruction to the prompt: ``Optimize the time complexity of your algorithm.''

As shown in Figure~\ref{fig:baseline_hint}, this simple optimization-oriented prompt substantially reduced the number of unsuccessful Java solutions across all models. The improvement was most pronounced on medium and hard problems, indicating that explicit optimization cues can effectively guide LLMs toward generating more efficient and correct code. These findings suggest that LLMs can respond adaptively to concise, goal-directed instructions, highlighting the potential of prompt engineering as a lightweight yet effective strategy for enhancing LLM performance on challenging programming tasks.

\subsection{RQ3: Code Correction with Iterative Feedback}\label{sec:result_rq3}

\begin{figure}[!t]
\centering
\includegraphics[width=0.95\linewidth]{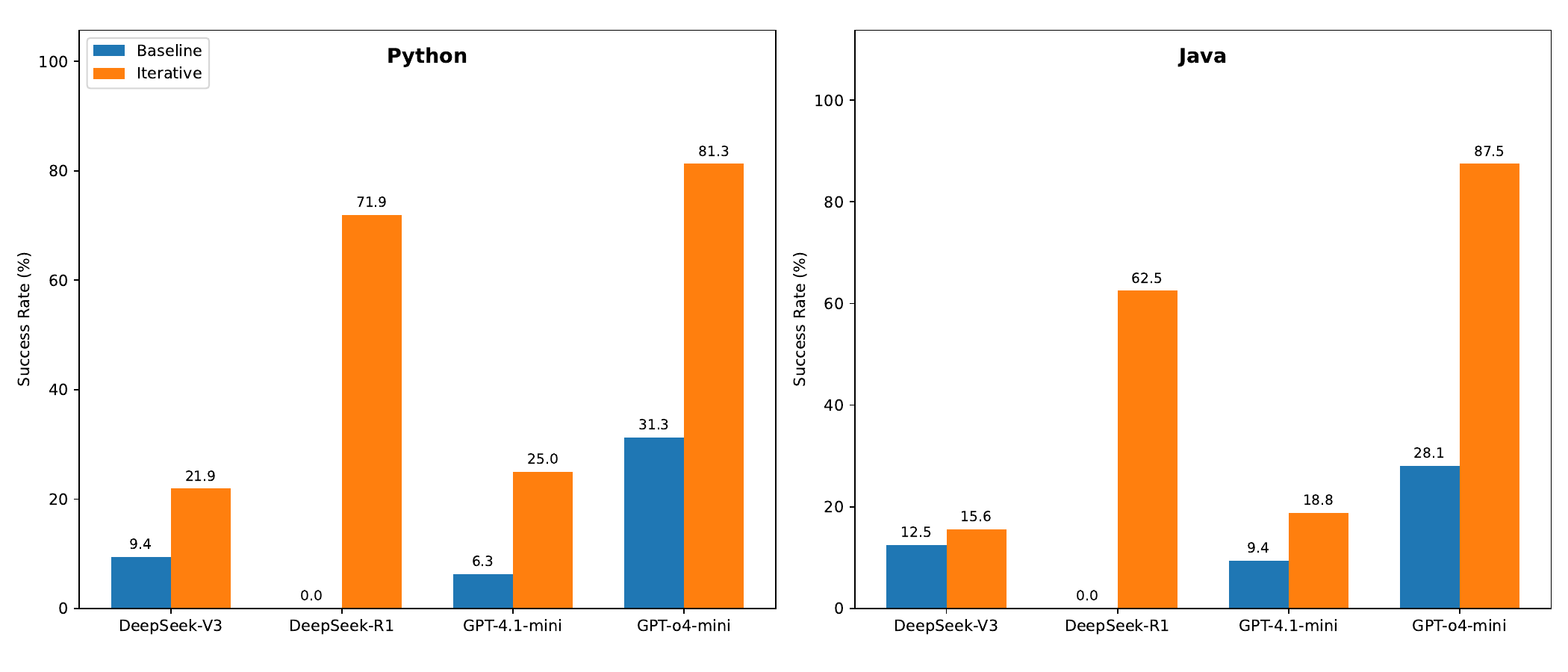}
\caption{Success rate comparison, baseline vs iterative framework.}
\label{fig:baseline_iter_compare}
\vspace{-0.1in}
\end{figure}

Building on the previous finding that instructive prompts can guide model behavior, we next investigated whether LLMs can autonomously improve through feedback-based iteration. Specifically, we examined how models respond when provided with informative feedback such as runtime error messages and failed test case details. To this end, we selected 32 of the most frequently failed problems across all models to construct the Challenge Dataset (Section~\ref{sec:datasets}) and applied the iterative refinement framework described in Section~\ref{sec:iterative_procedure} to four LLMs across two programming languages.

\begin{figure}[!t]
\centering
\includegraphics[width=0.8\linewidth]{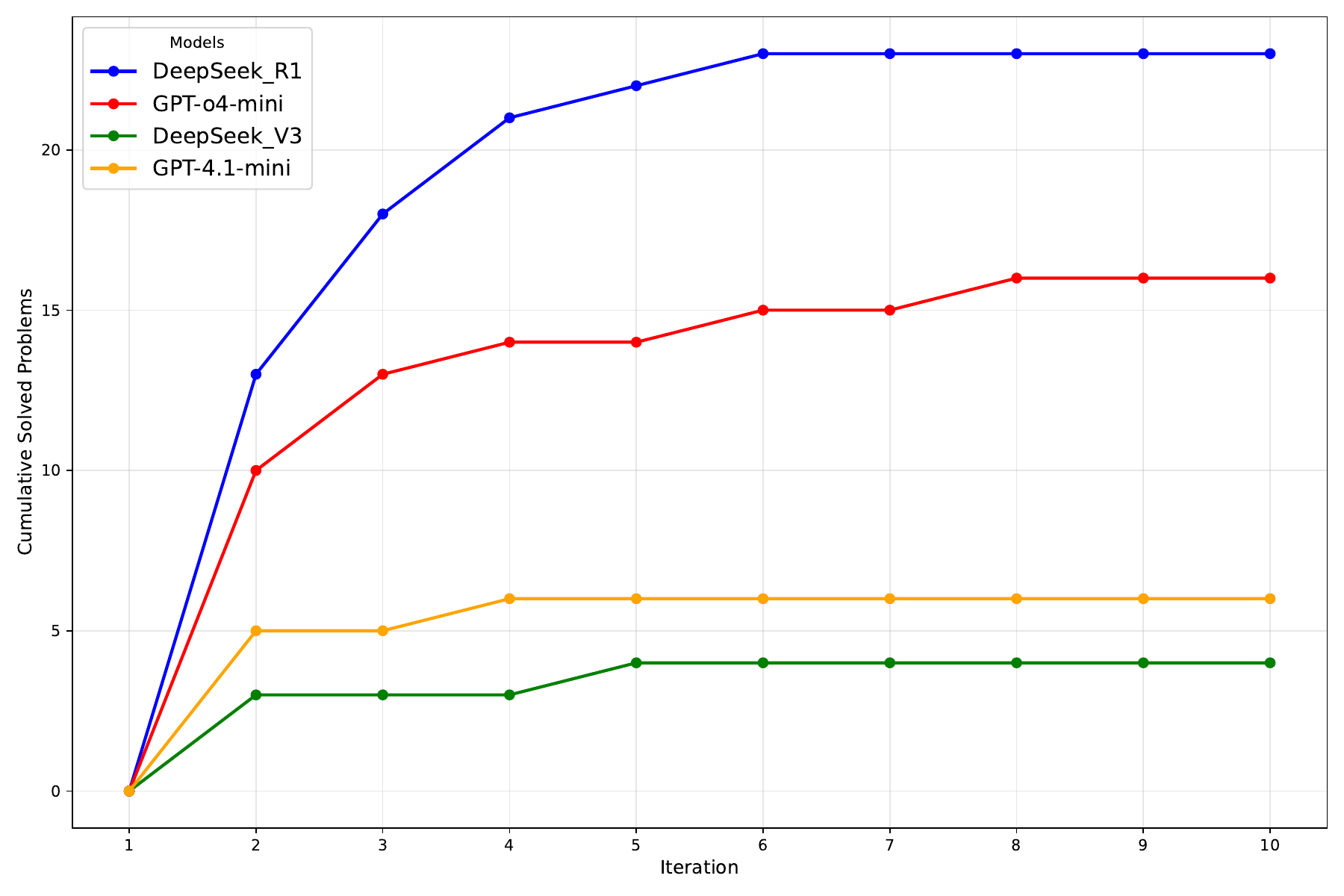}
\vspace{-0.1in}
\caption{Cumulative ISR@10 for Python solutions.}
\label{fig:iterative-isr-python}
\end{figure}

\begin{figure}[!t]
\vspace{-0.05in}
\centering
\includegraphics[width=0.82\linewidth]{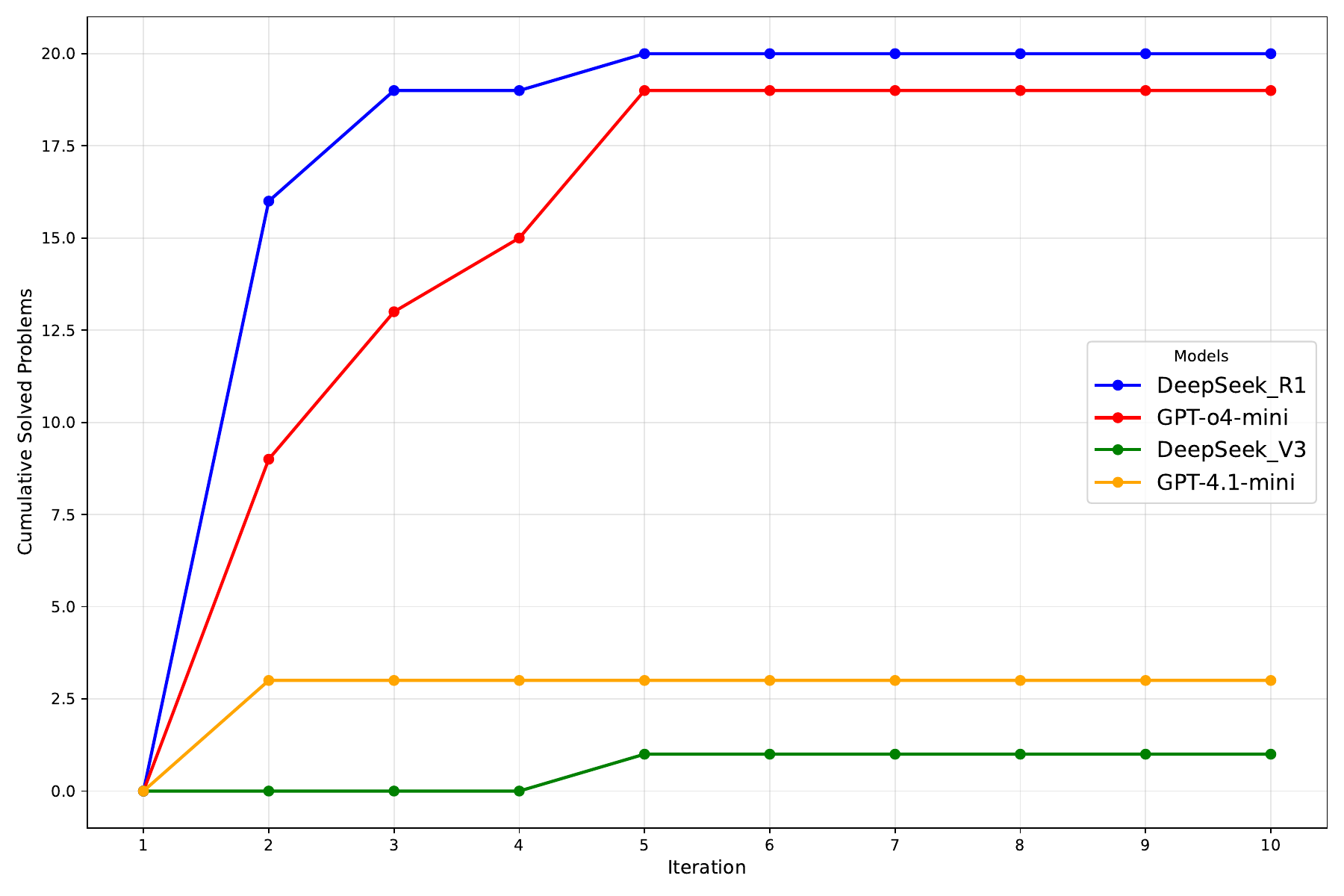}
\vspace{-0.1in}
\caption{Cumulative ISR@10 for Java solutions.}
\label{fig:iterative-isr-java}
\vspace{-0.1in}
\end{figure}

As shown in Figure~\ref{fig:baseline_iter_compare}, all models exhibited substantial gains in success rate compared to their single-attempt performance. Figures~\ref{fig:iterative-isr-python} and~\ref{fig:iterative-isr-java} further illustrate the cumulative number of successful solutions in Python and Java, respectively. DeepSeek-R1 and GPT-o4-mini, in particular, demonstrated consistent improvement across iterations, suggesting that these models can effectively interpret and act upon execution feedback. In contrast, DeepSeek-V3 and GPT-4.1-mini showed smaller but still noticeable gains, indicating limited but present feedback utilization capability.

Overall, these results suggest that LLMs can indeed leverage execution feedback to iteratively refine their code, though the degree of improvement varies by model and programming language. This finding highlights the potential of feedback-driven refinement as a powerful strategy for enhancing LLM problem-solving robustness beyond single-attempt code generation.

\subsection{RQ4: Reasoning vs. Non-Reasoning Models}\label{sec:result_rq4}

Figures~\ref{fig:iterative-isr-python} and~\ref{fig:iterative-isr-java} reveal a clear distinction between reasoning models (DeepSeek-R1 and GPT-o4-mini) and non-reasoning models (DeepSeek-V3 and GPT-4.1-mini). The reasoning models consistently improved their success rates across multiple iterations, showing steady gains as they incorporated execution feedback to refine their solutions. In contrast, the non-reasoning models exhibited only marginal improvements, often plateauing after one or two iterations.

These results indicate that models equipped with explicit reasoning capabilities, such as chain-of-thought (CoT) mechanisms, are more adept at interpreting and utilizing feedback signals during iterative refinement. By contrast, non-reasoning models, despite comparable architectures and model sizes, lack the internal reasoning structure needed to translate feedback into meaningful code revisions. This finding highlights that iterative improvement depends not only on model scale or training data, but also on the model's underlying reasoning capacity to generalize from feedback and self-correct effectively.

\begin{table}[t]
\caption{Statistic results for error analysis.}
\label{tab:error_analysis}
\begin{center}
\begin{small}
\begin{sc}
\resizebox{\textwidth}{!}{
\begin{tabular}{llccc@{\hspace{1em}}ccc@{\hspace{1em}}ccc@{\hspace{1em}}ccc@{\hspace{1em}}ccc}
\toprule
\multirow{2}{*}{Model} & \multirow{2}{*}{Language} & \multicolumn{3}{c}{\begin{tabular}{@{}c@{}}Compile\\ Error\end{tabular}} & \multicolumn{3}{c}{\begin{tabular}{@{}c@{}}Runtime\\ Error\end{tabular}} & \multicolumn{3}{c}{\begin{tabular}{@{}c@{}}Wrong\\ Answer\end{tabular}} & \multicolumn{3}{c}{\begin{tabular}{@{}c@{}}Time Limit\\ Exceeded\end{tabular}} & \multicolumn{3}{c}{\begin{tabular}{@{}c@{}}Memory Limit\\ Exceeded\end{tabular}} \\
\cmidrule(lr){3-5} \cmidrule(lr){6-8} \cmidrule(lr){9-11} \cmidrule(lr){12-14} \cmidrule(lr){15-17}
& & total & fixed & \% & total & fixed & \% & total & fixed & \% & total & fixed & \% & total & fixed & \% \\
\midrule
DeepSeek-R1 & Python & 0 & 0 & -- & 14 & 10 & 71.4 & 75 & 45 & 60.0 & 39 & 6 & 15.4 & 3 & 3 & 100.0 \\
& Java & 2 & 2 & 100.0 & 1 & 1 & 100.0 & 93 & 52 & 55.9 & 50 & 12 & 24.0 & 0 & 0 & -- \\
\midrule
GPT-o4-mini & Python & 0 & 0 & -- & 12 & 11 & 91.7 & 65 & 34 & 52.3 & 14 & 6 & 42.8 & 0 & 0 & -- \\
& Java & 9 & 8 & 88.9 & 3 & 2 & 66.7 & 59 & 32 & 54.2 & 8 & 1 & 12.5 & 0 & 0 & -- \\
\midrule
DeepSeek-V3 & Python & 0 & 0 & -- & 1 & 1 & 100.0 & 210 & 60 & 28.6 & 42 & 6 & 14.3 & 4 & 0 & 0.0 \\
& Java & 5 & 4 & 80.0 & 3 & 1 & 33.3 & 237 & 73 & 30.8 & 28 & 9 & 32.1 & 1 & 0 & 0.0 \\
\midrule
GPT-4.1-mini & Python & 0 & 0 & -- & 2 & 2 & 100.0 & 222 & 53 & 23.9 & 23 & 4 & 17.4 & 1 & 0 & 0.0 \\
& Java & 1 & 1 & 100.0 & 10 & 0 & 0.0 & 223 & 58 & 26.0 & 26 & 2 & 7.7 & 3 & 1 & 33.3 \\
\midrule
Overall & Python & 0 & 0 & -- & 29 & 24 & 82.8 & 572 & 192 & 33.6 & 118 & 22 & 18.6 & 8 & 3 & 37.5 \\
& Java & 17 & 15 & 88.2 & 17 & 4 & 23.6 & 612 & 215 & 35.1 & 112 & 24 & 21.4 & 4 & 1 & 25.0 \\
\bottomrule
\end{tabular}
}
\end{sc}
\end{small}
\end{center}
\vspace{0.2in}
\end{table}

\subsection{RQ5: Error Analysis}\label{sec:result_rq5}

In the iterative refinement experiments, we analyzed the distribution and correction rates of different error types based on LeetCode's standard categories: Compile Error, Runtime Error, Wrong Answer, Time Limit Exceeded, and Memory Limit Exceeded. The evaluation reports provided both the error type and the number of passed testcases for each iteration.

To formally characterize when an error was considered fixed, we adopt the following definition:

\begin{definition}
Let $T_i$ denote the number of testcases passed by the solution at iteration $i$.  
An error observed at iteration $i$ is considered \textbf{fixed} if and only if
\begin{equation}
T_{i+1} > T_i.
\end{equation}
In other words, a fix is recognized when the subsequent iteration yields an improvement in the number of successfully passed testcases.
\end{definition}

Using this definition, we derived aggregate statistics across all models, as summarized in Table~\ref{tab:error_analysis}. The majority of failures, approximately 95\%, belonged to the Wrong Answer and Time Limit Exceeded categories. These two types were also the most challenging to correct, with average fix rates of roughly 35\% and 20\%, respectively. The low fix rate for Time Limit Exceeded errors likely stems from their algorithmic nature, often involving inefficient logic, infinite loops, or suboptimal data structures, issues that require substantial code restructuring rather than minor syntax or logic edits.

Interestingly, Compile Error appeared exclusively in Java solutions, suggesting that Python's simpler and more flexible syntax makes it easier for LLMs to maintain syntactic correctness. Runtime Errors occurred more frequently in Python but were also among the easiest to correct, likely because they arise from direct execution failures that provide clear feedback signals.

Overall, these results indicate that syntactic and runtime errors are the most amenable to correction through iterative refinement, while algorithmic inefficiencies and logical errors remain considerably harder to resolve. This pattern underscores the current strength of LLMs in local debugging and incremental code repair, contrasted with their more limited capacity for deep algorithmic reasoning and structural optimization.

\section{Key Findings and Limitations}\label{sec:discussion}

This study provides a systematic analysis of LLMs' capabilities for iterative code refinement, moving beyond static, single-attempt evaluations to highlight their feedback-driven potential.

\vspace{-0.15in}
\subsection{Key Findings}

\subsubsection{Iterative evaluation reveals latent capabilities:} ISR@k demonstrates that many models can solve challenging problems through multiple attempts, highlighting adaptive potential missed by \textit{pass@1}.

\vspace{-0.15in}
\subsubsection{Reasoning capacity matters:} Reasoning models (DeepSeek-R1, GPT-o4-mini) steadily improve across iterations, while non-reasoning models plateau quickly, indicating that chain-of-thought abilities enhance feedback utilization.

\vspace{-0.15in}
\subsubsection{Prompt guidance enhances efficiency:} Optimization-oriented prompts reduced inefficiency errors, showing that instructive guidance—whether via prompts or integrated feedback—improves refinement outcomes.

\vspace{-0.15in}
\subsubsection{Error-type fixability varies:} Syntactic and runtime errors are easier to fix (often $>$80\%), whereas logical and algorithmic errors (Wrong Answer, Time Limit Exceeded) remain challenging ($<$35\%), revealing current limitations in deep algorithmic reasoning.

\vspace{-0.15in}
\subsubsection{Implications for Benchmarking:}
Relying solely on \textit{pass@1} risks misrepresenting a model's practical utility. Metrics like ISR@k and MIS better capture robustness, iterative improvement, and efficiency, providing a more realistic assessment aligned with real-world coding workflows.

\vspace{-0.1in}
\subsection{Limitations}

\subsubsection{Dataset scope:} Evaluation focused on LeetCode algorithmic problems, which may not generalize to domains like web development or systems programming.

\vspace{-0.15in}
\subsubsection{Structured feedback:} Feedback from LeetCode evaluation system was idealized and automated; real-world errors are often ambiguous or incomplete.

\vspace{-0.15in}
\subsubsection{Computational cost:} Iterative refinement is resource-intensive, highlighting the need for smarter iteration strategies.

\vspace{-0.15in}
\subsubsection{Potential regressions:} Repeated iterations may introduce new errors or vulnerabilities, which current metrics do not capture.
\section{Related Work}\label{sec:related_work}

Generative AI models have rapidly advanced in synthesizing source code from natural language, reshaping software development. Early models like Codex~\cite{chen2021evaluating} achieved syntactically and functionally correct programs on benchmarks such as HumanEval~\cite{chen2021evaluating}, while CodeGen~\cite{nijkamp2023codegen} extended capabilities to multi-turn code synthesis. Production systems like GitHub Copilot demonstrate the real-world applicability of these models~\cite{jiang2024survey}.

Standard evaluation metrics, including $pass@1$ and $pass@k$, assess single-attempt correctness but ignore iterative refinement, a key aspect of human programming~\cite{doderlein2025piloting,chen2021evaluating,jiang2023selfevolve,jiang2024survey,niu2024evaluating,dougherty2025proving,guimaraes2025analyzing,walder2025pass,masood2025,zhang2025holistic}. Recent studies address this limitation: G-Pass$@k$~\cite{liu2024your} measures consistency across generations, and Pass@k Training~\cite{chen2025pass} leverages pass-based metrics as optimization objectives.

A growing body of work explores LLM self-correction via feedback. Frameworks such as ORPS~\cite{yu2024outcome}, LDB~\cite{zhong2024debug}, and CRITIC~\cite{gou2023critic} show that incorporating verification and runtime feedback enhances performance. Systems like SelfEvolve~\cite{jiang2023selfevolve}, CodeIt~\cite{butt2024codeit}, VeriMind~\cite{nadimi2025verimind}, and GPU Kernel Scientist~\cite{andrews2025gpu} demonstrate iterative refinement across domains, highlighting models' ability to act as autonomous agents.

Beyond correctness, interaction design~\cite{shukla2025security} and prompt adaptation~\cite{pawar2024exploring,uusnakki2025exploring} improve alignment with user intent, as shown in ChatCoder~\cite{wang2023chatcoder} and Parsing Requirements for Automatic Prompting~\cite{davis2025parsing}. Surveys~\cite{jiang2024survey,wang2023review,niu2024evaluating} emphasize holistic evaluation encompassing adaptability, reasoning depth, and self-correction.

Despite these advances, gaps remain in systematically quantifying how LLMs leverage feedback to recover from errors across models, error types, and programming languages. This study addresses this challenge with a large-scale, multi-model investigation into iterative refinement framework, providing empirical insights into LLM reasoning and error recovery dynamics.

\vspace{-0.1in}
\section{Conclusion and Future Work}\label{sec:conclusion}

This study explored the iterative self-correction capabilities of LLMs in code generation through a novel iterative refinement framework that simulates real-world programming workflows. Experiments across three datasets, four models, and two languages showed that LLMs can improve success rates through iterative refinement, though gains vary with model architecture, reasoning ability, and error type. Reasoning models such as DeepSeek-R1 and GPT-o4-mini steadily improved across iterations, while non-reasoning models plateaued early. Syntactic and runtime errors were often resolved quickly, whereas logical and algorithmic errors remained challenging, highlighting current limits in deep algorithmic reasoning.

Future work should explore richer feedback mechanisms, like model-generated unit tests or symbolic execution traces, and extend iterative evaluation to other software engineering tasks such as vulnerability repair, code translation, and legacy system modernization. Additionally, fine-tuning models specifically for iterative debugging, potentially via reinforcement learning or hindsight replay, may further enhance reliability and efficiency in multi-step reasoning tasks.

By quantifying how LLMs learn from mistakes and adapt through feedback, this work not only advances the scientific understanding of the ability of LLMs but also offers practical metrics and methodologies for building next-generation, feedback-driven AI programming assistants.

\appendix
\vspace{.5in}
\section{Tables}

\begin{table}[H]
\caption{pass@1 scores across different models and datasets.}
\label{tab:dataset-performance}
\vskip 0.15in
\begin{center}
\begin{small}
\begin{sc}
\begin{tabular}{llcccc}
\toprule
\begin{tabular}{@{}c@{}}Dataset\end{tabular} & Language & DeepSeek-V3 & DeepSeek-R1 & GPT-4.1-mini & GPT-o4-mini \\
\midrule
\multirow{2}{*}{\begin{tabular}{@{}l@{}}Core\\dataset\end{tabular}} 
& Python & 72.4 & 84.0 & 76.2 & \textbf{89.1} \\
& Java & 71.6 & 82.4 & 75.1 & \textbf{87.3} \\
\midrule
\multirow{2}{*}{\begin{tabular}{@{}l@{}}Strain\\dataset\end{tabular}}
& Python & 55.5 & 79.0 & 67.0 & \textbf{87.0} \\
& Java & 55.5 & 76.5 & 66.5 & \textbf{84.0} \\
\midrule
\multirow{2}{*}{\begin{tabular}{@{}l@{}}Challenge\\dataset\end{tabular}}
& Python & 9.4 & 0.0 & 6.3 & \textbf{31.3} \\
& Java & 12.5 & 0.0 & 9.4 & \textbf{28.1} \\
\bottomrule
\end{tabular}
\end{sc}
\end{small}
\end{center}
\vskip -0.1in
\end{table}

\begin{table}[H]
\caption{Iterative metrics results for DeepSeek-R1 on various temperature and top-p settings.}
\label{tab:hyperparameter-tuning}
\vskip 0.15in
\begin{center}
\begin{small}
\begin{sc}
\begin{tabular}{l@{\hspace{.5em}}c@{\hspace{1em}}c@{\hspace{1em}}c@{\hspace{1em}}c@{\hspace{1em}}c@{\hspace{1em}}c}
\toprule
& \multicolumn{2}{c}{top-p=0.3} & \multicolumn{2}{c}{top-p=0.6} & \multicolumn{2}{c}{top-p=0.9} \\
\cmidrule(lr){2-3} \cmidrule(lr){4-5} \cmidrule(lr){6-7}
& ISR@10 & MIS & ISR@10 & MIS & ISR@10 & MIS \\
\midrule
T=0.1 & 62.5 & 9.0 & 62.5 & 7.5 & 56.2 & 9.0 \\
T=0.5 & 62.5 & 7.0 & 62.5 & 4.0 & 62.5 & 8.0 \\
T=0.9 & 68.8 & 4.0 & 68.8 & 6.5 & 62.5 & 4.0 \\
\bottomrule
\end{tabular}
\end{sc}
\end{small}
\end{center}
\vskip -0.1in
\end{table}

\section{Case Study}

Two representative cases are presented from this iterative refinement experiment to illustrate when LLMs successfully improve their solutions and when their capability remains limited.

\begin{table}[H]
\caption{Example of successful iterative refinement.}
\label{tab:case_good}
\vskip 0.15in
\begin{center}
\begin{small}
\begin{sc}
\begin{tabular}{lc@{\hspace{1em}}c@{\hspace{1em}}c}
\toprule
iteration & 1 & 2 & 3 \\
\midrule
Result & Wrong Answer & Wrong Answer & Accepted \\
\begin{tabular}{@{}c@{}}Testcase\\ Coverage\end{tabular} & 98.3\% & 99.1\% & 100.0\% \\
\bottomrule
\end{tabular}
\end{sc}
\end{small}
\end{center}
\vskip -0.1in
\end{table}

\subsection{Case \#1}

GPT-4o-mini’s improvement on the LeetCode problem “Minimum Cost to Equalize Array” is particularly noteworthy. As shown in Table \ref{tab:case_good}, although its initial attempt failed due to a wrong answer, the model effectively incorporated feedback over the following two iterations, progressively increasing testcase coverage and ultimately solving the problem by iteration \#3.

\begin{figure}[h]
\centering
\begin{minipage}[t]{0.48\textwidth}
\begin{lstlisting}[language=Python, frame=lines, basicstyle=\fontsize{8}{10}\ttfamily, lineskip=2pt, numbers=left]
# Python code at Iteration #1
...
candidates = {base}
if T0 >= base:
    candidates.add(T0)
    candidates.add(T0 + 1)
candidates.add(base + 1)
\end{lstlisting}
\end{minipage}
\hfill
\begin{minipage}[t]{0.48\textwidth}
\begin{lstlisting}[language=Python, frame=lines, basicstyle=\fontsize{8}{10}\ttfamily, lineskip=2pt, numbers=left]
# Python code at Iteration #2
...
n = len(nums)
candidates = {base, base + 1}
if n >= 3:
    candidates.add(T0i)
    candidates.add(T0i - 1)
\end{lstlisting}
\end{minipage}
\caption{GPT-o4-mini solutions comparison, iteration \#1 vs iteration \#2.}
\label{fig:case_good_compare}
\vspace{-0.1in}
\end{figure}

In the first attempt, the failed testcase was \verb|nums| $= [55, 52, 29, 11]$, \verb|cost1| $= 18$, \verb|cost2| $= 2$, where the expected output was 118 but the model produced 122. This error likely resulted from an unhandled corner case. As illustrated in Figure \ref{fig:case_good_compare}, the initial solution evaluated only two candidates, \verb|base| and \verb|(base + 1)|, and considered \verb|T0| and \verb|(T0 + 1)| only when \verb|T0| $\geq base$. In contrast, the revised solution evaluated all four candidates whenever \verb|len(nums)| $\geq 3$. In this failed testcase, where \verb|len(nums)| $= 4$, the expanded candidate set in the second solution enhanced corner-case coverage and successfully resolved the issue.

\begin{figure}[h]
\centering
\begin{minipage}[t]{0.48\textwidth}
\begin{lstlisting}[language=Python, frame=lines, basicstyle=\fontsize{8}{10}\ttfamily, lineskip=2pt, numbers=left]
# Python code at Iteration #2
...
n = len(nums)
candidates = {base, base + 1}
if n >= 3:
    candidates.add(T0i)
    candidates.add(T0i - 1)
\end{lstlisting}
\end{minipage}
\hfill
\begin{minipage}[t]{0.48\textwidth}
\begin{lstlisting}[language=Python, frame=lines, basicstyle=\fontsize{8}{10}\ttfamily, lineskip=2pt, numbers=left]
# Python code at Iteration #3
...
candidates = {base, base + 1,
                T0 - 1, T0, T0 + 1}
\end{lstlisting}
\end{minipage}
\caption{GPT-o4-mini solutions comparison, iteration \#2 vs iteration \#3.}
\label{fig:case_good_compare2}
\vspace{-0.1in}
\end{figure}

Although the second solution passed one previously failed testcase, it encountered another failure on \verb|nums| $= [60, 19, 53, 31, 57]$, \verb|cost1| $= 60$, \verb|cost2| $= 2$, where the expected output was 90 but the model produced 144. This error likely stemmed from insufficient candidate coverage. As noted earlier, the second solution considered four candidates, \verb|base|, \verb|(base + 1)|, \verb|T0i|, and \verb|(T0i - 1)|, which proved inadequate for this scenario. Consequently, GPT-4o-mini further refined its approach, as shown in Figure \ref{fig:case_good_compare2}. In this version, the model removed conditional checks and directly evaluated all five possible candidates, thereby expanding its search space and improving robustness against corner cases. This revision enabled the model to pass all testcases successfully.

This case exemplifies the LLM’s ability to iteratively refine its reasoning and self-correct functional errors, demonstrating clear progress in problem understanding and solution generalization through iterative feedback.

\subsection{Case \#2}

\begin{table}[H]
\caption{Example of unsuccessful iterative refinement.}
\label{tab:case_study2}
\vskip 0.15in
\begin{center}
\begin{small}
\begin{sc}
\resizebox{\textwidth}{!}{
\begin{tabular}{lc@{\hspace{.5em}}c@{\hspace{.5em}}c@{\hspace{.5em}}c@{\hspace{.5em}}c@{\hspace{.5em}}c@{\hspace{.5em}}c@{\hspace{.5em}}c@{\hspace{.5em}}c@{\hspace{.5em}}c}
\toprule
Iteration & 1 & 2 & 3 & 4 & 5 & 6 & 7 & 8 & 9 & 10 \\
\midrule
Result & TLE\footnotemark{} & TLE & TLE & TLE & TLE & TLE & TLE & TLE & TLE & TLE \\
\begin{tabular}{@{}c@{}}Testcase\\ Coverage\end{tabular} & 99.56\% & 99.56\% & 99.56\% & 99.56\% & 99.85\% & 99.56\% & 99.56\% & 99.56\% & 99.56\% & 99.56\% \\
\bottomrule
\end{tabular}
}
\end{sc}
\end{small}
\end{center}
\vskip -0.1in
\end{table}
\footnotetext{TLE stands for Time Limit Exceeded error.}

In this case study, we analyze the iterative refinement behavior of DeepSeek-R1 on the LeetCode problem “Longest Special Path II”. As summarized in Table~\ref{tab:case_study2}, all iterations consistently resulted in Time Limit Exceeded (TLE) outcomes, indicating that none of the revisions successfully optimized the algorithm to meet the efficiency constraints of the problem. 

\begin{figure}[h]
\centering
\begin{minipage}[t]{0.48\textwidth}
\begin{lstlisting}[language=Python, frame=lines, basicstyle=\fontsize{8}{10}\ttfamily, lineskip=2pt, numbers=left]
# Java code at Iteration #1
...
for (int i = savedGlobalLeft;
        i < globalLeft; i++) {
    int node = path.get(i);
    int vv = nums[node];
    freq[vv]++;
    ...
}
\end{lstlisting}
\end{minipage}
\hfill
\begin{minipage}[t]{0.48\textwidth}
\begin{lstlisting}[language=Python, frame=lines, basicstyle=\fontsize{8}{10}\ttfamily, lineskip=2pt, numbers=left]
# Java code at Iteration #6
...
for (int value : state.removals) {
    freq[value]++;
    ...
}
\end{lstlisting}
\end{minipage}
\caption{DeepSeek-R1 solutions comparison, iteration \#1 vs iteration \#6.}
\label{fig:case_bad_compare}
\vspace{-0.1in}
\end{figure}

If we examine the Java solutions generated by DeepSeek-R1 for this problem, we can observe in Figure \ref{fig:case_bad_compare} a notable \verb|globalLeft| restoration logic at the end of the DFS method (iteratoin \#1). This design has clear performance implications:

\begin{itemize}
\item The \verb|path| can be $O(n)$ long in the worst case (e.g., a linear chain tree).
\item \verb|globalLeft| may move $O(n)$ positions during a single DFS call.
\item The restoration loop runs $O(n)$ times for each DFS call.
\item DFS itself makes $O(n)$ calls in total.
\end{itemize}

Consequently, the worst-case time complexity of this algorithm is $O(n^2)$, which is inefficient.

In subsequent iterations, the model attempted to address this issue by adopting the approach shown on the right-hand side of Figure \ref{fig:case_bad_compare} (iteration \#6). However, in the worst case (a degenerate tree resembling a linked list), each node may accumulate up to $O(n)$ entries in its \verb|state.removals| list. Processing these entries requires $O(n)$ operations per node. For such a degenerate tree:

\begin{itemize}
\item There are $n$ nodes along the path.
\item Each node processes up to $n$ removals.
\end{itemize}

This results in a total complexity of $O(n^2)$, which is no better than the previous solution.

Despite repeated refinement attempts, the model’s testcase coverage remained nearly constant (around 99.56\%) across iterations. This suggests that DeepSeek-R1 made only superficial changes without addressing the core performance bottleneck. This example highlights a limitation in the model’s ability to iteratively improve solutions when fundamental algorithmic efficiency issues are involved.

\end{document}